\documentclass[aps,prl,twocolumn,showpacs]{revtex4}
\usepackage{graphicx}
\usepackage{amssymb}

\begin{document}

\title{Evidence for Jahn-Teller distortions at the antiferromagnetic transition in  LaTiO$_3$}

\author{J. Hemberger$^1$, H.-A. Krug von Nidda$^1$, V. Fritsch$^1$, J. Deisenhofer$^1$, S.~Lobina$^1$, %M. V. Eremin$^{4}$,
 T.~Rudolf$^1$, P.~Lunkenheimer$^1$, F.
Lichtenberg$^1$, A. Loidl$^1$, D.~Bruns$^2$, and
B.~B\"{u}chner$^2$}
 \affiliation{\mbox{$^1$ EKM, Institut f\"{u}r Physik, Universit\"{a}t Augsburg, D-86135 Augsburg, Germany }\\
 %$^4$ Kazan State University, 420008 Kazan, Russia \\
 %$^2$ II. Physikalisches Institut der Universit\"{a}t zu  K\"{o}ln, % %Z\"{u}lpicher Str. 77,
 %D-50937 K\"{o}ln, Germany \\
$^2$ II. Physikalisches Institut A, RWTH Aachen, D-52056 Aachen,
Germany}

\begin{abstract}
LaTiO$_3$ is known as Mott-insulator which orders
antiferromagnetically at $T_{\rm N}=146$~K. We report on results
of thermal expansion and temperature dependent x-ray diffraction
together with measurements of the heat capacity, electrical
transport measurements, and optical spectroscopy in untwinned
single crystals. At $T_{\rm N}$ significant structural changes
appear, which are volume conserving. Concomitant anomalies are
also observed in the dc-resistivity, in bulk modulus, and optical
reflectivity spectra. We interpret these experimental observations
as evidence of orbital order.

%lifting the ground state degeneracy of the d-electrons.

%At $T_N$ a significant anisotropic structural anomaly has been
%detected. This anomaly appears to be volume conserving and is
%compatible with the onset of an orbital order scenario.
%Corresponding anomalies are observed in measurements of the
%C-resistivity, the bulk modulus, and in optical reflectivity
%spectra.
\end{abstract}


\maketitle

In transition-metal oxides a variety of complex electronic ground
states is evoked by the interplay of spin, charge, orbital, and
lattice degrees of freedom \cite{review}. Amongst those, orbital
degeneracy and orbital ordering (OO) have to be considered as key
features to understand the physical properties of many compounds.
In principle, OO can be detected via neutron diffraction
\cite{Rodrig}, resonant x-ray scattering \cite{Murakami}, nuclear
magnetic resonance (NMR) \cite{ito99}, or electron spin resonance
\cite{Deisen} techniques. However, in most cases lattice
distortions induced by OO phenomena are detected, while the OO
parameter and the ground state wave functions remain hidden. In
this article we report on structural anomalies in LaTiO$_3$
observed at the antiferromagnetic (AFM) ordering temperature which
indicate the onset of OO induced by spin ordering. The perovskite
LaTiO$_3$, where Ti reveals a 3$d^1$ electronic configuration, can
be characterized as a Mott-Hubbard insulator \cite{oki95}
revealing a $G$-type AFM order below the N\'{e}el temperature
$T_{\rm N} = 146$~K \cite{lic91} with an ordered moment of $0.45
\mu_{\rm B}$ \cite{gor83}.

The orbital ground state in LaTiO$_3$ has not been determined
unambiguously: The quasi-cubic crystal field \cite{mei99} (CF)
splits the $3d$-levels into a low-lying $t_{2g}$ triplet and an
excited $e_g$ doublet. The complexity of the problem of a single
electron in a threefold degenerate $t_{2g}$ level, including
spin-orbit (SO) coupling and long-range spin order has been
outlined long ago by Goodenough \cite{Goo68} and by Kugel and
Khomskii \cite{kugelkhom}. Very recently the orbital ground state
of LaTiO$_3$ again came under heavy debate both theoretically
\cite{kha00,Kha01,kha02,moch01,kik03} and experimentally
\cite{ito99,kei00,fri02,Arao02,koeln}. Khaliullin and Maekawa
\cite{kha00} proposed the scenario of an orbital liquid to explain
the reduced magnetic moment and the small spin-wave anisotropy
observed in neutron-scattering experiments \cite{kei00}. They
predicted a gap of the orbital excitations, which however could
not be observed experimentally \cite{fri02}. To solve this
discrepancy a refined model has been proposed, which may explain
the dominant magnon contribution to the specific heat
\cite{kik03}.
% taking into account the
%interplay between spin and orbital degrees of freedom.
% This approach removes the orbital degeneracy
%and yields heavily damped orbital excitations, which in turn can
%explain the fact that the low-temperature heat capacity is
%dominated by magnon excitations alone.\cite{fri02}
Mochizuki and Imada \cite{moch01} investigated several scenarios
that lift the $t_{2g}$ degeneracy based on the competition of SO
coupling, Jahn-Teller (JT) effect, and CF due to GdFeO$_3$-type
lattice distortions for $R$TiO$_3$ with ($R=$ La, Sm, Nd, Gd, Y).
%suggested that in LaTiO$_3$ the $t_{2g}$ orbital degeneracy is
%lifted completely and that the orbital moment is fully quenched.
%Later-on, these authors included the effect of a CF caused by the
%La$^{3+}$ ions which lifts the degeneracy of the $3d$
%orbitals.\cite{moch01}
%They proposed that the GdFeO$_3$-type distortions characteristic
%for many $R$TiO$_3$ ($R=$ La, Y, $\dots$) perovskites are of
%paramount importance for the competition between spin-orbit
%interaction and Jahn-Teller (JT) effect.
While in YTiO$_3$ the JT distortion was shown to be dominant, the
influence of SO coupling must be considered important for
LaTiO$_3$  %, but still is not completely understood
\cite{moch01,kha02}. Also the experimental situation is not yet
settled. Itoh et al. \cite{ito99} have explained their NMR results
in LaTiO$_3$ assuming a degenerate orbital ground state. Neutron
scattering provided no experimental evidence for OO but has been
interpreted in terms of an orbital liquid \cite{kei00}. However,
an anisotropy of the magnetization was observed as well in the
paramagnetic (PM) as in the AFM regime in Ref.~\onlinecite{fri02}.
A small JT distortion has been derived from the observation of
atomic displacements by means of transmission electron microscopy
\cite{Arao02} and, utilizing X-ray and neutron diffraction, the
observation of structural anomalies at $T_{\rm N}$ was reported
recently \cite{koeln}.
% as evidence against an orbital liquid state

The purpose of this work is to give a detailed structural
characterization of LaTiO$_3$, including thermal expansion and
phonon properties and to relate these to electronic,
thermodynamic, and magnetic properties. The observed anomalies
%provide arguments for the onset
are interpreted in terms of OO.

Untwinned single crystals of LaTiO$_3$ have been prepared by
floating-zone melting as described elsewhere \cite{lic91}. The
oxygen content was determined by thermogravi\-metry. The x-ray
diffraction pattern at $T=295$~K revealed an orthorhombic
structure (P$bnm$, $z = 4$) with the lattice parameters
$a=5.633$~\AA,
$b=5.617$~\AA, $c=7.915$~\AA. %5.5968
%These lattice parameters are almost cubic and close to those
%reported in literature (e. g. Ref.~\cite{mei99}).
Laue measurements were performed on the single crystalline samples
to orient the samples and to exclude twinning.
%The orthorombic directions $a$, $b$, and $c$ could be identified
%from the comparison of temperature dependent x-ray results and
%thermal expansion measurements (see below).
%The magnetization measurements were performed with a commercial
%SQUID-system (MPMS, {\sc QuantumDesign}) between 1.8~K and 400~K.
The electrical resistivity was measured with a four-probe
electrometer circuit in the temperature range $30 \leq T \leq
300$~K. The specific heat was obtained with noncommercial setups
utilizing a quasi-adiabatic method between 2~K and 15~K and an
AC-method between 10~K and 300~K. X-ray powder diffraction was
performed in the temperature range $90 \leq T \leq 350$~K
employing a
{\sc Stoe} diffractometer %({\sc Stadi-P})
with a nitrogen gas-flow cryostat. The linear thermal expansion
coefficient $\alpha= (\partial L/
\partial T) /L$ was determined utilizing a home-built high-resolution
capacitance dilatometer. The measurements of the optical
reflectivity were carried out on polished single crystals using a
Fourier transform IR spectrometer ({\sc Bruker IFS 113v}).

\begin{figure}[tb]
\includegraphics[clip,width=75mm]{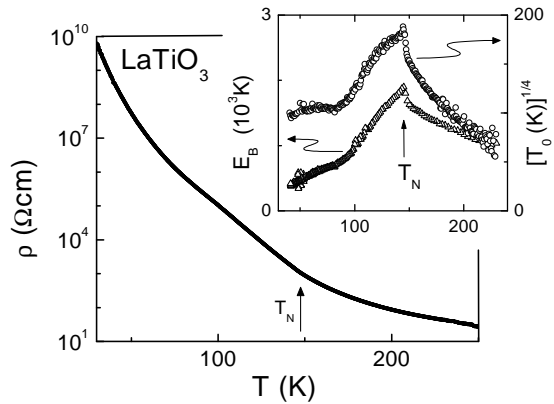}
\caption{DC-resistivity of LaTiO$_3$. The inset shows the $T$
dependence of the effective energy barrier assuming an
Arrhenius-type behavior $\rho=\rho_0 \exp (E_{\rm B}/T)$ (left
scale) or the characteristic temperature of 3-dimensional VRH
$\rho\propto\exp [(T_0/T)^{1/4}]$ (right scale) as described in
the text.\label{rho}}
\end{figure}

Figure~\ref{rho} displays measurements of the DC-resistivity. At
all temperatures, above and below $T_{\rm N}$, the resistivity
$\rho(T)$ exhibits semiconducting characteristics. The arrow
marks the magnetic transition at $T_{\rm N} =146$~K determined
from susceptibility measurements \cite{fri02} for the sample under
investigation.
%In this representation only a weak anomaly can be
%detected.
It is worth mentioning that this temperature corresponds to the
highest transition temperatures reported in literature
\cite{lic91,mei99}. It is known that LaTiO$_3$ is sensitive to
deviations from its nominal stoichiometry
\cite{lic91,oka93,tag99,cra94,fri01}, driving the
Mott-Hubbard-insulator away from half filling towards a metallic
state yielding a concomitant strong reduction of $T_{\rm N}$.
Thus, the high resistivity and the high transition temperature
show that the samples under investigation are close to the nominal
composition. The inset of Fig.~\ref{rho} gives a representation
according to the assumption of thermally activated transport and
variable range hopping (VRH), respectively. The derivatives
$E_B=$d(ln$\rho)/$d$(1/T)$ (left scale) and
$T_0^{1/4}=$d(ln$\rho)/$d$(1/T^{1/4})$ (right scale) are plotted
versus temperature. These quantities can be interpreted as
generalized energy barrier $E_B$ in the case of purely activated
transport or as characteristic temperature $T_0$ assuming VRH.
Both curves reveal a strong variation from far above to well below
the magnetic ordering transition with a spike-like enhancement
just at $T_{\rm N}$. From the inset in Fig.~\ref{rho} it is clear
that neither a purely thermally activated behavior nor
3-dimensional (3D) hopping transport can account for the $T$
dependence of the resistivity in LaTiO$_3$. At this point we
propose that not only spin-dependent scattering processes are
responsible for the anomaly of the resistivity around $T_{\rm N}$,
but rather that orbital fluctuations have to be taken into
account. It is reasonable to assume that the inter-site transfer
matrix elements, responsible for the charge transport, could
strongly be influenced by the orbital dynamics. It is also
interesting to note that below 100~K, 3D-VRH dominates the charge
transport, demonstrated by the temperature independent value of
the respective derivative shown in the inset of Fig.~\ref{rho}.
For temperatures $T \leq 100$~K the derivative approaches a
constant value with $T_0 \approx 10^8$~K, which is close to those
values observed in doped manganites \cite{par00}, where OO is
established. %indicating the fixed OO.

\begin{figure}[tb]
\includegraphics[clip,width=75mm]{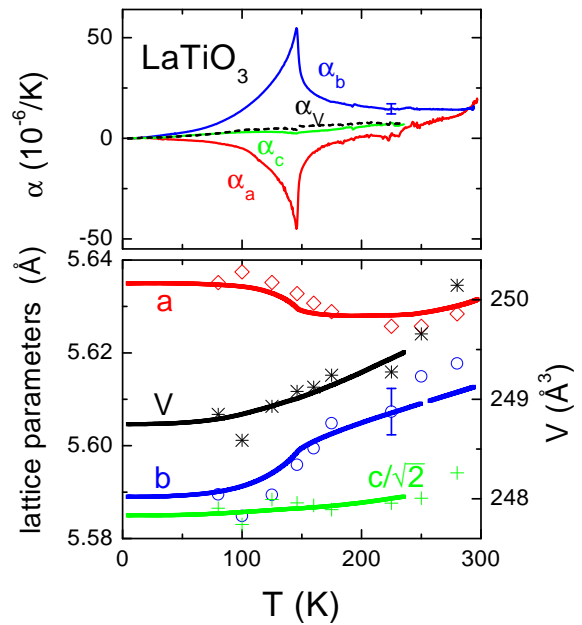}
\caption{Upper frame: Thermal expansion coefficients of LaTiO$_3$
for the three different crystallographic directions. The dashed
line gives the averaged volume expansion coefficient
$\alpha_V=(\alpha_a+\alpha_b+\alpha_c)/3$. Lower frame: $T$
dependence of the lattice parameters $a$, $b$, $c/\sqrt{2}$ (left
scale), and cell volume $V$ (right scale). The symbols denote data
obtained by x-ray diffraction. The solid lines were calculated by
the integration of $\alpha$ and scaled to the lattice constants.
The error bar given in the lower frame refers only to the
uncertainty of the absolute values gained by %x-ray diffraction and
the scaling procedure. The relative resolution is by far higher.
\label{fig2}}
\end{figure}

Looking for further indications of OO close to $T_{\rm N}$, we
investigated the $T$ dependence of the linear thermal expansion
coefficient $\alpha=(\partial L/
\partial T) /L$ for LaTiO$_3$. The results for the different
crystallographic directions are shown in the upper frame of
Fig.~\ref{fig2}. At the AFM transition a sharp peak-like anomaly
can be detected along $a$ and $b$ direction. In contrast, the
magnetic transition can hardly be detected in the measurements
along $c$. The thermal expansion coefficient along the $a$- and
$b$-direction reaches values up to $\left| {\alpha}\right| =
5\times 10^{-5}$ K$^{-1}$, about ten times higher than e.g. those
near the displacive phase transition in non-magnetic SrTiO$_3$ at
107~K \cite{wil76}. But both directions $a$ and $b$ have opposite
sign: while $\alpha_b$ is positive for all temperatures,
$\alpha_a$ is negative for $T < 200$~K. Obviously, the crystal
expands along the $a$-direction with decreasing temperature. At
the same time, the averaged volume thermal expansion coefficient
$\alpha_V=(\alpha_a+\alpha_b+\alpha_c)/3$ (dashed line in
Fig.~\ref{fig2}) exhibits only a weak increase with temperature as
expected for ordinary solids and no anomaly can be found within
the uncertainty of the measurement, i.e. the structural anomaly at
$T_{\rm N}$ is volume conserving. It is remarkable, that the
deviations of $\alpha_a$ and $\alpha_b$ from $\alpha_V$ set in
already well above $T_{\rm N}$. While the measurements of the
thermal expansion allow a much higher relative resolution compared
to diffraction methods, it is difficult to derive absolute values
of the lattice parameters from the dilatation data. For this
reason we combined these results with temperature dependent x-ray
diffraction measurements down to $T=70$~K. The results for the $T$
dependence of the lattice constants (left scale) and the volume of
the unit cell (right scale) are displayed in the lower frame of
Fig.~\ref{fig2}. Again it is remarkable that the cell volume
$V(T)$ reveals no anomaly at $T_{\rm N}$. The same is valid for
$c(T)$. The lattice parameters $a(T)$ and $b(T)$ exhibit distinct
anomalies with opposite signs. However, even though the difference
between $a$ and $b$ is strongly increased and $b$ nearly equals
$c/\sqrt{2}$ for low temperatures, the orthorhombic crystal
symmetry is not changed. We interpret these phenomena as an
isostructural orbital order-disorder transition resulting from OO
similar to the case of LaMnO$_3$ \cite{Sanchez03,gra99}.

In perovskites the orthorhombic $O^{\prime}$ phase with
$a>c/\sqrt{2}$ points towards a cooperative JT distortion.
Although the lattice parameters of LaTiO$_3$ at room temperature
already fulfill this condition, to the best of our knowledge the
JT effect has never been considered in the PM regime in this
compound, but is very likely to appear in the magnetically ordered
phase: Following Goodenough \cite{Goo68}, the observed distortions
around $T_N$ can be understood in terms of the energy-level scheme
constructed from the $t_{2g}$ states which resemble p-like
electron states (effective orbital moment $\tilde{L}=1$). SO
coupling and the low symmetry component of the CF split this
orbital triplet, which is occupied by one electron (spin $S=1/2$),
into three Kramers doublets even above $T_N$. In the perovskite
structure magnetic order is usually found to initiate the JT
effect in the $t_{2g}$ levels \cite{Goo68}, whereas this is not
necessarily the case for the $e_g$ levels like for example in
LaMnO$_3$ ($T_{\rm N} = 140$~K), where the cooperative JT effect
persists up to 750~K and at $T_N$ only an additional enhancement
of spin-phonon coupling can be observed \cite{gra99}. Hence, in
LaTiO$_3$, we interpret the observed structural anomalies at
$T_{\rm N}$ in terms of a cooperative JT distortion as expected
for the $t_{2g}$ levels just below the magnetic transition in
magnets with collinear spin order. The underlying OO can be
understood taking into account the $p$-wave symmetry of the
$\tilde{L}=1$ manifold. A $p$-like orbital induces a uniaxial
elongation of a TiO$_6$ octahedron along two opposite Ti-O bonds.
In the ground-state the $p$-like orbitals are fixed along this
O-Ti-O direction within the $ab$ plane and order in a
ferrodistorsive pattern induced by the $G$-type AFM spin order.
This amplifies the weak orthorhombic distortion, which (possibly
together with orbital fluctuations) is already present in the PM
phase, resulting in the experimentally observed expansion along
the crystallographic $a$ direction.

\begin{figure}[tb]
\includegraphics[clip,width=70mm]{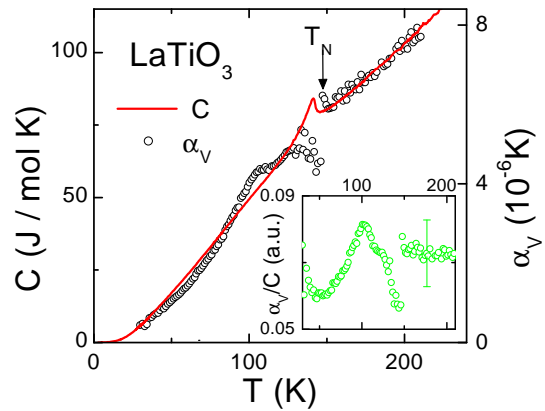}
\caption{Specific heat (left scale) and volume thermal expansion
coefficient (right scale) of LaTiO$_3$. Lower inset: Volume
Gr\"uneisenparameter $\Gamma_V \propto \alpha_V/C$ averaged from
the three crystallographic directions devided by the bulk modulus
$B$.\label{gruen}}
\end{figure}

The $T$ dependence of the specific heat $C$ of LaTiO$_3$ is
displayed in Fig.~\ref{gruen} together with the averaged thermal
expansion coefficient. The behavior of $C(T)$ coincides very well
with that of $\alpha_V$ above the magnetic transition, but
exhibits deviations at and below $T_{\rm N}$. These deviations can
better be illustrated in terms of the thermodynamic
Gr\"uneisenparameter $\Gamma = 3 \alpha_V V B/C$. In the inset of
Fig.~\ref{gruen} we show the $T$ dependence of the ratio of the
volume thermal expansion coefficient and the heat capacity,
$\alpha_V V / C \propto \Gamma/B$. In normal anharmonic crystals
$\Gamma$ is only weakly temperature dependent and the bulk modulus
should slightly increase with decreasing temperature. However, it
is also well known that the acoustic phonons directly couple to
the orbital (quadrupolar) degrees of freedom and, hence, the bulk
modulus should be sensitive to orbital fluctuations and to the
onset of OO. One example of the softening of longitudinal acoustic
modes at the JT transition in doped manganites can be found in
Ref.~\onlinecite{par00}. For $T > T_{\rm N}$, the ratio $\Gamma/B
\propto \alpha_V V / C$ is nearly constant. It reveals a distinct
and strong anomaly at $T_{\rm N}$, is strongly enhanced just below
the ordering temperature and subsequently passes through a minimum
on further decreasing temperature. The minimum at low temperatures
can be explained by magnetic excitations \cite{fri02}, which will
strongly affect the specific heat, but much less the thermal
expansion.
%Magnon contributions to
%the heat capacity have been reported in Ref.~\onlinecite{fri02}.
It seems straightforward to explain the maximum in $\Gamma/B$
just below $T_{\rm N}$ by the onset of OO. As has been documented
in the manganites, the longitudinal acoustic modes, and hence
also the bulk modulus, reveal a significant softening at the JT
transition \cite{par00}.

\begin{figure}[tb]
\includegraphics[clip,width=85mm]{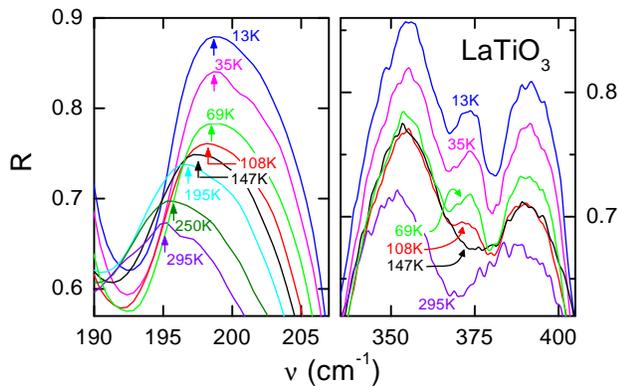}
\caption{Spectra of the optical reflectivity for various
temperatures. The data illustrates the $T$ dependence of selected
phonon modes. \label{optik}}
\end{figure}

Finally we will show that in the AFM phase also the phonon
spectrum is strongly influenced, indicating significant changes of
the local structure and the binding energies. At room temperature
the reflectance of LaTiO$_3$ shows the three characteristic bands
of phonons of the perovskite structure grouped  around 200, 400,
and 600 cm$^{-1}$, which roughly can be ascribed to external,
bending, and stretching modes \cite{cra94}. In cubic symmetry only
three modes are IR active. In the orthorhombic P$nma$ structure
these bands split into single phonon excitations and
%according to group theory
25 infrared-active modes have to be expected \cite{may00}.
Fig.~\ref{optik} shows the reflectance in LaTiO$_3$ for two
selected groups of modes close to 200 and 400~cm$^{-1}$ for
various temperatures. A detailed report on the optical properties
will be given in a forthcoming paper \cite{rudolf}. The phonons at
200~cm$^{-1}$ belong to the external modes were the TiO$_6$
octahedra vibrate against the La ions. The modes close to
400~cm$^{-1}$ can be characterized as Ti-O bond-angle modulations.
Without going into further details, already the raw data provide
striking experimental evidence that close to $T_{\rm N}$ the
external modes reveal a significant shift of the eigenfrequencies
(indicated by arrows). At $T_{\rm N}$ these modes considerably
stiffen on decreasing temperatures. In the group of bending modes,
below $T_{\rm N}$ one mode significantly increases in the spectra
at 375~cm$^{-1}$ indicating that the local symmetry at $T_{\rm N}$
changes considerably.
%resulting in a strong redistribution of the optical weight.

In conclusion, we have presented detailed measurements of the
electrical resistivity, the $T$ dependence of lattice constants,
thermal expansion, and heat capacity, as well as of the optical
reflectivity for selected groups of phonons in stoichiometric
single crystalline and untwinned LaTiO$_3$. All results exhibit
anomalies slightly above or just at the magnetic ordering
temperature indicating significant structural changes. At this
transition the PM and orthorhombic crystal characterized by
lattice constants $a>b>c/\sqrt{2}$ transforms into an
antiferromagnet with almost tetragonal symmetry $(a>b\approx
c/\sqrt{2})$. We interpret these significant structural changes as
evidence of OO. Both the PM and the AFM phase can be consistently
described assuming a splitting of the $t_{2g}$ state into three
Kramers doublets. Below $T_N$ the $G$-type AFM order induces an
additional cooperative JT distortion, which results in a
ferrodistortive OO pattern. Even so it can not be ruled out, if
additional quantum fluctuations \cite{kha02} have to be taken into
account to describe the orbital groundstate, it is difficult to
explain such type of structural distortion related to the onset of
magnetic order within a picture of entirely obliterated orbital
order as proposed in terms of the orbital-liquid scenarios.

This work was supported by the BMBF via VDI/EKM, FKZ 13N6917/18
and by the DFG within SFB 484 (Augsburg). We are grateful to
M.~V.~Eremin for useful discussions.

%  ####################################
% \bibliographystyle{prlbst}
% \bibliography{latio3,labatio3}

\end{document}